\documentclass[12pt,a4paper,final]{iopart}
\usepackage{iopams}  
\usepackage{graphicx}
\usepackage{comment}
\usepackage{bm}

\expandafter\let\csname equation*\endcsname\relax
\expandafter\let\csname endequation*\endcsname\relax

\usepackage{amsmath}
\usepackage[utf8]{inputenc}

\usepackage[breaklinks=true,colorlinks=true,linkcolor=blue,urlcolor=blue,citecolor=blue]{hyperref}

\newcommand{\defeq}{\mathrel{\mathop:}=}
\newcommand{\dplus}[1]{\Delta^{+} #1}
\newcommand{\dminus}[1]{\Delta^{-} #1}
\newcommand{\dcenter}[1]{\Delta #1}
\newcommand{\indic}[1]{\mathbb{Q}_{#1}}
\renewcommand{\H}{\mathcal{H}}

\begin{document}

\title{Temperature and equipartition in discrete systems}

\author[cor1]{Sergio Davis$^{1,2}$}
\address{$^1$Research Center in the Intersection of Plasma Physics, Matter and Complexity (P$^2$mc), Comisi\'on Chilena de Energía Nuclear, Casilla 188-D, Santiago, Chile}
\address{$^2$Departamento de F\'isica, Facultad de Ciencias Exactas, Universidad Andres Bello. Sazi\'e 2212, piso 7, 8370136, Santiago, Chile.}
\ead{sergio.davis@cchen.cl}

\begin{abstract}
The generalized equipartition theorem known as the conjugate variables theorem (Phys. Rev. E \textbf{86}, 051136 [2012]), originally obtained in the context of statistical inference of 
continuous random variables, is extended in this work to the case of discrete variables. Using this new set of theorems we derive novel thermodynamic identities for the canonical ensemble 
connecting temperature with measurable observables.
\end{abstract}

\section{Introduction}

Temperature is defined in the microcanonical ensemble through the derivative
\begin{equation}
\label{eq:temp}
\frac{1}{T(E)} \defeq \frac{\partial \mathcal{S}(E)}{\partial E}
\end{equation}
where
\begin{equation}
\mathcal{S}(E) \defeq k_B \ln \Omega(E)
\end{equation}
is Boltzmann's definition of thermodynamic entropy~\cite{Greiner2012}. However, in the case of systems with a discrete set of energy levels this derivative is no longer properly defined, and 
a measurable temperature using \eqref{eq:temp} is only recovered in the thermodynamic limit, where the discrete levels become a continuum. An alternative, equivalent definition of 
temperature uses generalized equipartition theorems~\cite{Jepps2000} of the form
\begin{equation}
\label{eq:equipart}
\left<\frac{\partial \omega}{\partial \Gamma_k}\right>_\beta = \beta\left<\omega\frac{\partial \H}{\partial \Gamma_k}\right>_\beta
\end{equation}
in the canonical ensemble, where we also need to take the thermodynamic limit of a discrete system in order to properly handle the derivatives of the Hamiltonian on the right-hand side.
Here $\H(\bm \Gamma)$ is the Hamiltonian of the system, $\Gamma_k$ is any of the phase-space coordinates and $\omega = \omega(\bm \Gamma)$ is an arbitrary, differentiable phase-space function.

In this work we will extend the family of identities in \eqref{eq:equipart} to the case of discrete variables using forward and backward differences.

\section{Brief review of the conjugate variables theorem}

The conjugate variables theorem (CVT)~\cite{Davis2012, Davis2016c} is the expectation identity
\begin{equation}
\label{eq:cvt}
\big<\partial\omega\big>_S + \big<\omega\partial \ln p\big>_S = 0
\end{equation}
where $\partial$ is the partial derivative of a function with respect to its first argument, $\omega(X)$ is an arbitrary, differentiable function of $X$ and
\begin{equation}
p(X) \defeq P(X|S)
\end{equation}
is the probability density of the continuous random variable $X$. This form of the CVT requires that $P(X|S)$ vanishes at its boundaries, otherwise there is an additional 
boundary term that must be added to the right-hand side. In the more general case, if $X \in [a, b]$ then the CVT takes the form
\begin{equation}
\label{eq:cvt_boundaries}
\big<\partial\omega\big>_{S, B} + \big<\omega\partial \ln p\big>_{S, B} = p(b)\omega(b)-p(a)\omega(a)
\end{equation}
where $B$ represents the proposition $X \in [a, b]$ imposing the boundary. More details about this derivation can be found in \ref{sec:appendix}.

\section{Forward and backward versions of the discrete CVT}

In this section we will derive a generalization of \eqref{eq:cvt} for discrete variables where the partial derivatives $\partial$ will be replaced by finite differences.
Consider a discrete variable $n \in \{a, a+1, \ldots, b-1, b\}$ with probabilities
\begin{equation}
p_n \defeq P(n|S)
\end{equation}
and such that the expectation of a quantity $A_n$ is given by
\begin{equation}
\big<A\big>_S \defeq \sum_{n=a}^b p_n A_n.
\end{equation}

\noindent
Let us now define the forward difference by
\begin{equation}
\big(\dplus{A}\big)_n \defeq A_{n+1}-A_n
\end{equation}
and the backward difference by
\begin{equation}
\big(\dminus{A}\big)_n \defeq A_n-A_{n-1}.
\end{equation}

Using these definitions, we can rewrite the expectation of the forward difference $\big<\dplus{\omega}\big>_S$ with $\omega_n$ an arbitrary quantity, as
\begin{equation}
\begin{split}
\big<\dplus{\omega}\big>_S & = \sum_{n=a}^b p_n \big(\omega_{n+1}-\omega_n\big) \\
& = \sum_{n=a+1}^{b+1} p_{n-1}\omega_n - \sum_{n=a}^b p_n\omega_n \\
& = p_b\omega_{b+1} - \sum_{n=a}^b p_n\left[\frac{\big(\dminus{p}\big)_n}{p_n}\right]\omega_n.
\end{split}
\end{equation}

\noindent
We have then established the forward CVT, namely the identity
\begin{equation}
\label{eq:forward_cvt}
\big<\dplus{\omega}\big>_S + \left<\omega\left[\frac{\dminus{p}}{p}\right]\right>_S = p_b\,\omega_{b+1}.
\end{equation}

Note that this forward CVT only involves an upper boundary term. Similarly, we can start with the expectation of the backward difference $\big<\dminus{\omega}\big>_S$ and obtain
\begin{equation}
\begin{split}
\big<\dminus{\omega}\big>_S & = \sum_{n=a}^b p_n \big(\omega_n - \omega_{n-1}\big) \\
& = \sum_{n=a}^b p_n\omega_n - \sum_{n=a-1}^{b-1} p_{n+1}\omega_n \\
& = -p_a\omega_{a-1} - \sum_{n=a}^b p_n\left[\frac{\big(\dplus{p}\big)_n}{p_n}\right]\omega_n.
\end{split}
\end{equation}

\noindent
Therefore the backward CVT is the identity
\begin{equation}
\label{eq:backward_cvt}
\big<\dminus{\omega}\big>_S + \left<\omega\left[\frac{\dplus{p}}{p}\right]\right>_S = -p_a\omega_{a-1},
\end{equation}
in which only a lower boundary term appears. By defining the centered difference
\begin{equation}
\big(\dcenter{A}\big)_n \defeq \big(\dplus{A}\big)_n + \big(\dminus{A}\big)_n  = A_{n+1} - A_{n-1}
\end{equation}
and summing \eqref{eq:forward_cvt} and \eqref{eq:backward_cvt}, we see that
\begin{equation}
\big<\dcenter{\omega}\big>_S + \left<\omega\left[\frac{\dcenter{p}}{p}\right]\right>_S = p_b\,\omega_{b+1}-p_a\omega_{a-1},
\end{equation}
and it is clear that we recover \eqref{eq:cvt_boundaries} in the continuous limit.

\section{Examples in statistics}

\subsection{The Poisson distribution}

Consider a variable $k \geq 0$ that follows a Poisson distribution,
\begin{equation}
P(k|\lambda) = \frac{\exp(-\lambda)\lambda^k}{k!}.
\end{equation}

\noindent
Now, because
\begin{equation}
\lim_{k \rightarrow \infty} P(k|\lambda) = 0
\end{equation}
it makes sense to use the forward CVT, therefore we first need to compute
\begin{equation}
\frac{P(k-1|\lambda)}{P(k|\lambda)} = \frac{\lambda^{k-1}}{(k-1)!}\frac{k!}{\lambda^k} = \frac{k}{\lambda}.
\end{equation}

\noindent
The identity in \eqref{eq:forward_cvt} simply becomes
\begin{equation}
\big<\dplus{\omega}\big>_\lambda + \left<\omega\left[1-\frac{k}{\lambda}\right]\right>_\lambda = 0,
\end{equation}
and using $\omega_k = 1$ we have
\begin{equation}
1 - \frac{\big<k\big>_\lambda}{\lambda} = 0,
\end{equation}
recovering the well-known mean of a Poisson variable, $\big<k\big>_\lambda = \lambda$. Additionally, from the choice $\omega_k = k$ we get
\begin{equation}
1 + \big<k\big>_\lambda - \frac{\big<k^2\big>_\lambda}{\lambda} = 0
\end{equation}
then $\big<k^2\big>_\lambda = \lambda^2 + \lambda$ and the variance $\big<(\delta k)^2\big>_\lambda = \lambda$ is recovered.

\subsection{The binomial distribution}

Consider a variable $0 \leq n \leq N$ following a binomial distribution,
\begin{equation}
P(n|N, p) = \frac{N!}{n!(N-n)!}\,p^n(1-p)^{N-n}.
\end{equation}

\noindent
Here we will use the backward CVT, so we need to compute
\begin{equation}
\frac{P(n+1|N , p)}{P(n|N, p)} = \frac{p^{n+1}(1-p)^{N-n-1}}{(n+1)!(N-n-1)!}\frac{n!(N-n)!}{p^n(1-p)^{N-n}} = \frac{N-n}{n+1}\frac{p}{1-p}.
\end{equation}

\noindent
Replacing this into \eqref{eq:backward_cvt}, we have
\begin{equation}
\Big<\dminus{\omega}\Big>_{N, p} + \left<\omega\left[\frac{N-n}{n+1}\frac{p}{1-p}-1\right]\right>_{N, p} = -(1-p)^N\,\omega_{-1},
\end{equation}
and from the choice $\omega_n = n+1$ we obtain
\begin{equation}
\frac{p}{1-p}\Big(N-\big<n\big>_{N, p}\Big) = \big<n\big>_{N, p},
\end{equation}
hence the mean of a binomial variable is obtained as
\begin{equation}
\big<n\big>_{N, p} = p\,N.
\end{equation}

\section{Measurement of temperature in discrete canonical systems}

Now we will consider a system in canonical equilibrium having inverse temperature $\beta$. We will assume there is a discrete spectrum of energies $E_0, E_1, E_2, \ldots$ and the degeneracy 
of the $n$-th energy level is $\Omega_n$. Then, the canonical distribution of energies is
\begin{equation}
P(E_n|\beta) = \frac{\exp(-\beta E_n)\Omega_n}{Z(\beta)}
\end{equation}
with $Z(\beta)$ the partition function, given by
\begin{equation}
Z(\beta) = \sum_{n = 0}\exp(-\beta E_n)\Omega_n.
\end{equation}

\noindent
Because $n$ starts from zero upwards, we will use the backward CVT in \eqref{eq:backward_cvt},
\begin{equation}
\label{eq:temp_estim}
\Big<\dminus{\omega}\Big>_\beta + \left<\omega\left[\exp\left(-\beta\dplus{E}\right)\frac{\Omega_{n+1}}{\Omega_n} -1\right]\right>_\beta = -\frac{\exp(-\beta E_0)}{Z(\beta)}\,\Omega_0\,\omega_{-1}.
\end{equation}

Note that the value of $\Omega_0$ is not actually relevant, as the ratio $Z(\beta)/\Omega_0$ can be expressed in terms of a sum that only involves the relative degeneracies, namely
\begin{equation}
\frac{Z(\beta)}{\Omega_0} = \sum_{n = 0}\exp(-\beta E_n)\,\left(\frac{\Omega_n}{\Omega_0}\right).
\end{equation}

\subsection{Two level systems}

As an example, we will consider a system of $N$ two-level components, such that the $n$-th energy level of the entire system is given by
\begin{equation}
E_n = n\varepsilon_1 + (N-n)\varepsilon_0 = N\varepsilon_0 + n(\varepsilon_1-\varepsilon_0),
\end{equation}
where $\varepsilon_0$ and $\varepsilon_1$ are the energies of the ``down'' and ``up'' levels, respectively, and $n$ is the number of components in the ``up'' level. The degeneracy of the 
$n$-th level is
\begin{equation}
\Omega_n = \binom{N}{n} = \frac{N!}{n!(N-n)!}
\end{equation}
with $\Omega_0$ = 1. Moreover, we have
\begin{equation}
\dplus{E_n} = \varepsilon_1-\varepsilon_0,
\end{equation}
and
\begin{equation}
\frac{\Omega_{n+1}}{\Omega_n} = \frac{N-n}{n+1},
\end{equation}
so replacing these, we can simplify \eqref{eq:temp_estim} to
\begin{equation}
\Big<\dminus{\omega}\Big>_\beta = \left<\omega\left[1-\exp\left(-\beta(\varepsilon_1-\varepsilon_0)\right)\frac{N-n}{n+1}\right]\right>_\beta,
\end{equation}
provided we choose $\omega_{-1}$ to be zero. For instance, choosing $\omega_n = \Theta(n)(n+1)$ we have
\begin{equation}
\dminus\Big[\Theta(n)(n+1)\Big] = \Theta(n)(n+1)-\Theta(n-1)n = 1
\end{equation}
for $n \geq 0$. Therefore $\big<\dminus{\omega}\big>_\beta = 1$ and we can determine, after some algebra,
\begin{equation}
\beta = \frac{1}{\varepsilon_1-\varepsilon_0}\ln \left[\frac{N-\big<n\big>_\beta}{\big<n\big>_\beta}\right].
\end{equation}

\noindent
From this result we can readily deduce the caloric curve,
\begin{equation}
\label{eq:caloric}
\big<E\big>_\beta = N\varepsilon_0 + \frac{N(\varepsilon_1-\varepsilon_0)}{1+\exp\big(\beta(\varepsilon_1-\varepsilon_0)\big)},
\end{equation}
which reproduces the known result for two-level systems~\cite{Kardar2007}. Remarkably, we have been able to obtain the correct thermodynamics of the system without computing the partition function, 
or any sum over $n$. We can verify this result by explicit computation of the $n$-component partition function,
\begin{equation}
Z(\beta) = \Big[\exp(-\beta\varepsilon_0)+\exp(-\beta\varepsilon_1)\Big]^N = \exp(-\beta N\varepsilon_0)\Big[1+\exp\big(-\beta(\varepsilon_1-\varepsilon_0)\big)\Big]^N,
\end{equation}
which then gives
\begin{equation}
\big<E\big>_\beta = -\frac{\partial}{\partial \beta}\ln Z(\beta) = N\varepsilon_0 + N(\varepsilon_1-\varepsilon_0)\cdot\frac{\exp(-\beta(\varepsilon_1-\varepsilon_0))}{1+\exp(-\beta(\varepsilon_1-\varepsilon_0))},
\end{equation}
in agreement with \eqref{eq:caloric}.

\subsection{Quadratic number of states}

The second example involves a system with energy levels
\begin{equation}
E_n = n\varepsilon
\end{equation}
for $n \geq 0$, and degeneracy
\begin{equation}
\Omega_n = A n^2.
\end{equation}

As $\Omega_0$ = 0, the probability $P(E_0|\beta)$ also vanishes and we can take $n \geq 1$ for the expected values. Replacing $E_n$ and $\Omega_n$ into \eqref{eq:temp_estim} we simply get
\begin{equation}
\label{eq:cvt_ex2}
\Big<\dminus{\omega}\Big>_\beta = \left<\omega\left[1-\exp\left(-\beta\varepsilon\right)\Big(\frac{n+1}{n}\Big)^2\right]\right>_\beta.
\end{equation}

\noindent
Now, using the choice
\begin{equation}
\omega_n = n^2
\end{equation}
we obtain
\begin{equation}
\label{eq:dminus_ex2}
\dminus{(n^2)} = n^2 - (n-1)^2 = 2n -1
\end{equation}
for $n \geq 1$, and \eqref{eq:cvt_ex2} yields
\begin{equation}
2\big<n\big>_\beta - 1 = \big<n^2\big>_\beta - \exp(-\beta\varepsilon)\big<(n+1)^2\big>_\beta.
\end{equation}

\noindent
Solving for $\beta$ we finally obtain an inverse temperature estimator, given by
\begin{equation}
\label{eq:beta_ex2}
\beta = \frac{1}{\varepsilon}\ln\;\left[\frac{\big<(n+1)^2\big>_\beta}{\big<(n-1)^2\big>_\beta}\right].
\end{equation}

\noindent
We can verify this result using the property
\begin{equation}
\sum_{n=0}^\infty n^2\exp(-zn) = \frac{\partial^2}{\partial z^2}\left[\sum_{n=0}^\infty \exp(-zn)\right] = \frac{\partial^2}{\partial z^2}\left[\frac{\exp(z)}{\exp(z)-1}\right]
= \frac{1}{8}\frac{\text{sinh}(z)}{\text{sinh}(z/2)^4}
\end{equation}
with $z =  \beta\varepsilon$, through which we can directly obtain the partition function,
\begin{equation}
Z(\beta) = A\sum_{n=0}^\infty n^2\exp(-\beta \varepsilon n) = \frac{A}{8}\;\frac{\text{sinh}(\beta\varepsilon)}{\text{sinh}(\beta\varepsilon/2)^4}.
\end{equation}

\noindent
The first and second moments of $n$ can be obtained from the partition function as
\begin{subequations}
\begin{align}
\big<n\big>_\beta & = -\frac{1}{\varepsilon}\frac{\partial}{\partial \beta}\ln Z(\beta) = \frac{2+\text{cosh}(\beta\varepsilon)}{\text{sinh}(\beta\varepsilon)} \\
\big<n^2\big>_\beta & = \frac{1}{\varepsilon^2 Z(\beta)}\frac{\partial^2 Z(\beta)}{\partial \beta^2} = 1 + \frac{3}{\text{sinh}(\beta\varepsilon/2)^2},
\end{align}
\end{subequations}
respectively, so after some calculation we can verify that \eqref{eq:beta_ex2} holds.

\subsection{Three spins with Ising interaction}

\begin{table}
\begin{center}
\begin{tabular}{|c|c|c|c|c|}
\hline
$n$ & $s_1$ & $s_2$ & $s_3$ & $E_n$ \\
\hline
0 & -1 & -1 & -1 & $-3J$ \\
1 & 1 & 1 & 1 & $-3J$  \\
\hline
2 & -1 & -1 & 1 & $J$ \\
3 & -1 & 1 & -1 & $J$  \\
4 & -1 & 1 & 1 & $J$   \\
5 & 1 & -1 & -1 & $J$   \\
6 & 1 & -1 & 1 & $J$  \\
7 & 1 & 1 & -1 & $J$  \\
\hline
\end{tabular}
\end{center}
\caption{States and energies for a system of three Ising spins.}
\label{tbl:ising}
\end{table}

Now we will employ a different strategy than the one previously illustrated, this time without the use of the density of states. Consider a system of three spins interacting according to the Ising model, 
such that the Hamiltonian is
\begin{equation}
\H(s_1, s_2, s_3) = -J\big(s_1 s_2 + s_2 s_3 + s_1 s_3\big),
\end{equation}
with $s_i \in \{-1, 1\}$. There are 8 possible states of the system, having energies given in Table~\ref{tbl:ising}. We then use the canonical distribution of microstates directly, namely
\begin{equation}
P(n|\beta) = \frac{\exp(-\beta E_n)}{Z(\beta)},
\end{equation}
with $n$ an integer variable such that $0 \leq n \leq 7$, and construct its backward-CVT according to \eqref{eq:backward_cvt}. For this we first need to compute
\begin{equation}
\frac{P(n+1|\beta)}{P(n|\beta)} = \exp\Big(-\beta \big(\dplus{E}\big)_n\Big),
\end{equation}
and we obtain
\begin{equation}
\label{eq:cvt_ex3}
\big<\dminus{\omega}\big>_\beta = \Big<\omega\left[1-\exp\big(-\beta \dplus{E}\big)\right]\Big>_\beta,
\end{equation}
provided we choose $\omega_n$ such that $\omega_{-1} = 0$. This holds for the choice
\begin{equation}
\omega_n = \big(1-\delta(n, 7)\big)(n+1)
\end{equation}
and we can write
\begin{equation}
\label{eq:dminus_omega_ex3}
\big(\dminus{\omega}\big)_n = \big(1-\delta(n, 7)\big)(n+1) - \big(1-\delta(n-1, 7)\big)n = 1 - 8\delta(n,7).
\end{equation}

\noindent
By looking at Table~\ref{tbl:ising}, we obtain
\begin{equation}
\label{eq:dplus_E_ex3}
\big(\dplus{E}\big)_n = E_{n+1}-E_n = 4J\,\delta(n, 1)
\end{equation}
for $n < 7$. Replacing \eqref{eq:dminus_omega_ex3} and \eqref{eq:dplus_E_ex3} into \eqref{eq:cvt_ex3}, we have
\begin{equation}
\begin{split}
1 - 8\big<\delta(n, 7)\big>_\beta & = \left<\big(1-\delta(n, 7)\big)(n+1)\Big[1-\exp\big(-4\beta J\delta(n,1)\big)\Big]\right>_\beta \\
& = \left<(n+1)\big(1-\delta(n, 7)\big)\delta(n, 1)\right>_\beta \Big[1-\exp\big(-4\beta J\big)\Big] \\
& = 2\left<\big(1-\delta(n, 7)\big)\delta(n, 1)\right>_\beta \Big[1-\exp\big(-4\beta J\big)\Big] \\
& = 2\left<\delta(n, 1)\right>_\beta \Big[1-\exp\big(-4\beta J\big)\Big]
\end{split}
\end{equation}
where we have used the relation
\begin{equation}
\exp\big(-4\beta J\delta(n, 1)\big) = \delta(n, 1)\Big[\exp(-4\beta J)-1\Big] + 1
\end{equation}
and the fact that $\delta(n, 7)\delta(n, 1)$ = 0. After some algebra, we obtain an inverse temperature estimator in terms of two microstate probabilities, namely
\begin{equation}
\beta = -\frac{1}{4J}\ln \left[1-\frac{1 - 8\,p(++-)}{2\,p(+++)}\right]
\end{equation}
with $p(+++) \defeq P(n = 1|\beta)$ and $p(++-) \defeq P(n = 7|\beta)$. We can verify this result by using
\begin{align}
P(n = 1|\beta) & = \frac{\exp(3\beta J)}{Z(\beta)}, \\
P(n = 7|\beta) & = \frac{\exp(-\beta J)}{Z(\beta)},
\end{align}
and solving for the partition function, thereby obtaining the correct result,
\begin{equation}
Z(\beta) = 2\exp(3\beta J) + 6\exp(-\beta J).
\end{equation}

\section{Concluding remarks}

We have presented two expectation identities for discrete variables, namely the forward and backward versions of the conjugate variables theorem. In the case of the canonical ensemble, 
we have illustrated their use by two examples: a system of two-level components and a system with quadratic degeneracy, obtaining exact inverse temperature estimators in both cases.

\section{Acknowledgments}

The author gratefully acknowledges funding from ANID FONDECYT 1220651 grant.

\appendix

\section{Proof of the CVT with boundaries}
\label{sec:appendix}

We extend the variable $X \in [a, b]$ to $X \in \mathbb{R}$, such that
\begin{equation}
\int_a^b dx\,p(x) = 1,
\end{equation}
while the condition
\begin{equation}
P(X < a|S) = P(X > b|S) = 0
\end{equation}
still holds. A choice of probability density $\tilde{p}(x)$ that achieves this, is given by
\begin{equation}
\label{eq:ptilde}
\tilde{p}(x) \defeq p(x)\Theta(b-x)\Theta(x-a).
\end{equation}

\noindent
Letting $B$ be the proposition $a \leq X \leq b$ we recognize the indicator function
\begin{equation}
\label{eq:ptilde_indic}
\indic{B}(X) = \Theta(b-X)\Theta(X-a)
\end{equation}
and we note that $\tilde{p}(x)$ is actually the conditional probability density
\begin{equation}
P(X|S, B) = \frac{P(X|S)P(B|X)}{P(B|S)} = \tilde{p}(X)
\end{equation}
because $P(B|X) = \big<\indic{B}\big>_X = \indic{B}(X)$ and
\begin{equation}
P(B|S) = \int_{-\infty}^{\infty} dx P(X = x|S)\Theta(b-x)\Theta(x-a) = 1.
\end{equation}

The new density $P(X|S, B)$ fulfills all the conditions for the validity of the usual CVT, therefore we can directly write
\begin{equation}
\big<\partial \omega\big>_{S, B} + \big<\omega\partial \ln \tilde{p}\big>_{S, B}  = 0,
\end{equation}
which upon replacing \eqref{eq:ptilde} and \eqref{eq:ptilde_indic}, becomes
\begin{equation}
\big<\partial \omega\big>_{S, B} + \big<\omega\partial \ln p\big>_{S, B} = -\big<\omega\partial \indic{B}\big>_{S, B}.
\end{equation}

\noindent
As the derivative of the indicator function $\indic{B}$ reduces to
\begin{equation}
\frac{\partial \indic{B}(X)}{\partial X} = \delta(X-a)\Theta(b-X) - \delta(X-b)\Theta(X-a) = \delta(X-a)-\delta(X-b),
\end{equation}
we finally obtain
\begin{equation}
\big<\partial \omega\big>_{S, B} + \big<\omega\partial \ln p\big>_{S, B} = \Big<\omega\big[\delta(X-b)-\delta(X-a)\big]\Big>_{S, B}
= p(b)\omega(b) - p(a)\omega(a).
\end{equation}

\section*{References}
\bibliography{discrete}

\begin{thebibliography}{1}

\bibitem{Greiner2012}
W.~Greiner, L.~Neise, and H.~St{\"o}cker.
\newblock {\em Thermodynamics and statistical mechanics}.
\newblock Springer Science \& Business Media, 2012.

\bibitem{Jepps2000}
O.~G. Jepps, G.~S. Ayton, and D.~J. Evans.
\newblock Microscopic expressions for the thermodynamic temperature.
\newblock {\em Phys. Rev. E}, 62:4757--63, 2000.

\bibitem{Davis2012}
S.~Davis and G.~Guti\'errez.
\newblock Conjugate variables in continuous maximum-entropy inference.
\newblock {\em Phys. Rev. E}, 86:051136, 2012.

\bibitem{Davis2016c}
S.~Davis and G.~Gutiérrez.
\newblock Applications of the divergence theorem in {B}ayesian inference and
  {M}ax{E}nt.
\newblock {\em AIP Conf. Proc.}, 1757:20002, 2016.

\bibitem{Kardar2007}
M.~Kardar.
\newblock {\em Statistical Physics of Particles}.
\newblock Cambridge University Press, 2007.

\end{thebibliography}
\bibliographystyle{unsrt}

\end{document}